\documentclass[10pt,conference]{IEEEtran}

\usepackage{etoolbox}
\makeatletter
\def\ps@headings{%
	\def\@oddhead{\mbox{}\scriptsize\rightmark \hfil \thepage}%
	\def\@evenhead{\scriptsize\thepage \hfil \leftmark\mbox{}}%
	\def\@oddfoot{}%
	\def\@evenfoot{}}
\makeatother
\usepackage{verbatim}
\usepackage{xcolor}
\usepackage{amsfonts}
\usepackage{mathrsfs}
\usepackage{amsfonts}
\usepackage{amssymb}
\usepackage{graphicx}
\usepackage{epsfig}
\usepackage{epstopdf}
\usepackage{psfrag}
\usepackage{amsmath}
\usepackage{array}
\usepackage{cases}
\usepackage{subfigure}
\usepackage{cite,graphicx,amsmath,amssymb,color}
\usepackage{algorithmic}
\usepackage{algorithm}

\usepackage{algorithm}
\usepackage{algorithmic}
\usepackage{stmaryrd}
\usepackage{multirow}
\usepackage{subfig}
\usepackage{graphicx,times,amsmath} 

\usepackage{url}
\usepackage{enumerate}

\IEEEoverridecommandlockouts

\newtheorem{lemma}{Lemma}

\newtheorem{Exam}{Example}

\newtheorem{problem}{Problem}

\IEEEoverridecommandlockouts
\begin{document}
\bibliographystyle{IEEEtran}

\title{Optimal Multi-Quality Multicast for 360 Virtual Reality Video }
\author{\IEEEauthorblockN{Kaixuan Long, Chencheng Ye, Ying Cui}\IEEEauthorblockA{Shanghai Jiao Tong University, China}\and\IEEEauthorblockN{Zhi Liu}\IEEEauthorblockA{Shizuoka University, Japan}\thanks{The work of Y. Cui was supported by NSFC grant 61401272 and grant 61521062. The work of Z. Liu was supported by JSPS KAKENHI Grant JP16H02817 and JP18K18036 as well as National Institute of Informatics (NII) open collaborative research fund FY2018.}}
\maketitle

\begin{abstract}
A 360 virtual reality (VR) video, recording a scene of interest in every direction, provides VR users with immersive viewing experience. However, transmission of a 360 VR video which is of a much larger size than a traditional video to mobile users brings a heavy burden to a wireless network. In this paper, we consider multi-quality multicast of a 360 VR video from a single server to multiple users using time division multiple access (TDMA). To improve transmission efficiency, tiling is adopted, and each tile is pre-encoded into multiple representations with different qualities. We optimize the quality level selection, transmission time allocation and transmission power allocation to maximize the total utility of all users under the transmission time and power allocation constraints as well as the quality smoothness constraints for mixed-quality tiles. The problem is a challenging mixed discrete-continuous optimization problem. We propose two low-complexity algorithms to obtain two suboptimal solutions, using continuous relaxation and DC programming, respectively. Finally, numerical results demonstrate the advantage of the proposed solutions.
\end{abstract}


\begin{figure*}[!ht]
\normalsize{
\centering
\subfigure{
\begin{minipage}{16cm}
\centering
\includegraphics[width=16cm]{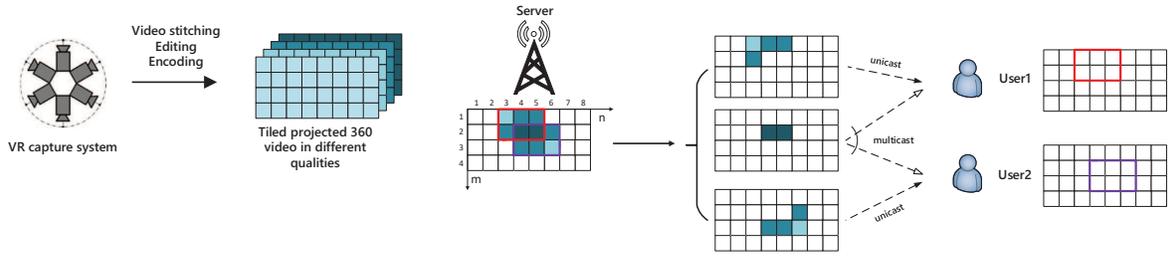}
\end{minipage}
}
\caption{  System model of multi-quality multicast of 360 VR video for 2 users.  }
\label{system}
}
\end{figure*}

\section{Introduction}\label{section_1}

A 360 \textit{virtual reality} (VR) video is generated by capturing a scene of interest in every direction at the same time using omnidirectional cameras.
A user wearing a VR headset (or Head Mounted Display (HMD)) can freely watch the scene of interest in any viewing direction at any time, hence enjoying immersing viewing experience. VR has vast applications in entertainment, education, medicine, etc.~\cite{8428401}. It is predicted that the global market of VR related products will reach 30 billion USD by 2020~\cite{111}.
A 360 VR video is of a much larger size than a traditional video. Thus, transmitting an entire 360 VR video brings a heavy burden to wireless networks. At any moment, a user watching a 360 VR video is interested in only one viewing direction, corresponding to one part of the 360 VR video, referred to as \textit{field-of-view} (FoV). Tiling is a technique proposed to enable flexible transmission of FoVs and improve transmission efficiency of a 360 VR video~\cite{8428401,7501810,Ju:2017:UWV:3097895.3097899,DBLP:journals/corr/El-Ganainy17,7532381}.
Specifically, a 360 VR video is divided into smaller rectangular segments of the same size, referred to as tiles, and any FoV can be composed by a set of tiles.
A VR user currently watching one FoV may switch to another FoV in a short time. To avoid view switch delay, the set of tiles covering a user's current FoV and the FoVs that may be watched shortly should
be transmitted simultaneously.

Considering heterogenous channel conditions and limited transmission resource, \cite{inoue2010interactive,Xie:2017:IQV:3123266.3123291,DBLP:journals/corr/GhoshAQ17,Ahmadi:2017:AMS:3126686.3126743} pre-encode each tile into multiple representations with different quality levels and consider quality adaptation in 360 VR video transmission.
Specifically, \cite{inoue2010interactive,Xie:2017:IQV:3123266.3123291,DBLP:journals/corr/GhoshAQ17} consider multi-quality 360 VR video transmission in single-user wireless networks. The proposed solutions in ~\cite{inoue2010interactive,Xie:2017:IQV:3123266.3123291,DBLP:journals/corr/GhoshAQ17} do not imply efficient designs for multi-user wireless networks. As far as we know, \cite{Ahmadi:2017:AMS:3126686.3126743} is the only work that considers multi-quality 360 VR video transmission in multi-user wireless networks and exploits multicast opportunities to improve transmission efficiency. In particular,~\cite{Ahmadi:2017:AMS:3126686.3126743} optimizes the quality level of each transmitted tile to maximize the total utility of all users, which reflects quality of experience (QoE) of all users. The size of the optimization problem is unnecessarily large, as the modulation and coding scheme (reflecting the quality level) of each tile is optimized separately, and the proposed heuristic algorithm may not provide desirable performance and complexity. In addition, the quality levels of adjacent tiles may vary significantly, leading to poor viewing experience.

In this paper, we would like to address the above limitations in multi-quality multicast for a 360 VR video in a wireless network with a single server and multiple users.
Specifically, we divide the 360 VR video into tiles and pre-encode each tile into multiple representations with different qualities using HEVC or H.264. For each user, we deliver a set of tiles that cover the user's current FoV and the FoVs that may be watched shortly.
We partition the set of tiles to be transmitted to users into subsets and multicast different subsets of tiles to different groups of users using Time Division Multiple Access (TDMA). We optimize the quality level selection, transmission time allocation and transmission power allocation for each subset of tiles to maximize the total utility of all users under the transmission time and power allocation constraints as well as the quality smoothness constraints for mixed-quality tiles.
Note that considering the optimization for each subset of tiles instead of each tile can reduce complexity significantly, without loss of optimality. In addition, note that considering the quality smoothness constraints can effectively avoid large quality variations for adjacent tiles, hence improving viewing experience.
The problem is a challenging mixed discrete-continuous optimization problem. We propose two low-complexity algorithms to obtain two suboptimal solutions. Specifically, the first suboptimal solution is obtained by transforming the continuous relaxation of the original problem into a convex problem and rounding the optimal solution of the convex problem. The second suboptimal solution is obtained by transforming the original problem into a Difference of Convex (DC) programming problem and providing a stationary point using a DC algorithm \cite{Lipp2016}. The second suboptimal solution achieves a higher total utility with a higher computational complexity than the first suboptimal solution.
Finally, numerical results demonstrate the advantage of the proposed suboptimal solutions. Note that this work extends our previous work~\cite{8428401} on optimal single-quality multicast for 360 VR video in TDMA systems. To the best of our knowledge, this is the first work providing low-complexity optimization-based multi-quality multicast for 360 VR video with quality smoothness guarantee.


\section{system model}\label{section_2}

As illustrated in Fig. \ref{system}, we consider downlink transmission of a 360 VR video from a single-antenna server (e.g., base station or access point) to $K$ $(\geq1)$ single-antenna users. Let $\mathcal{K}\triangleq\{1,...,K\}$ denote the set of user indices. Suppose the locations of all users do not change (in the considered timeframe). A user wearing a VR headset may be interested in one viewing direction, i.e., the center of one rectangular part of the 360 VR video, referred to as FoV, at sometime, and freely turn to another viewing direction after a while. The horizontal and vertical angular spans of each FoV are denoted by $F_{\text{h}}$ and $F_{\text{v}}$.

We consider tiling to enable flexible transmission of necessary FoVs and improve transmission efficiency of the 360 VR video.
Specifically, the 360 VR video is divided into $M\times N$ rectangular segments of the same size, referred to as tiles, where $M$ and $N$ represent the numbers of segments in each column and each row, respectively. The $(m,n)$-th tile refers to the tile in the $m$-th row and $n$-th column. Define $ \mathcal{M}\triangleq\{1,...,M\}$ and $\mathcal{N}\triangleq\{1,...,N\}$. Let $V_{\text{h}}$ and $V_{\text{v}}$ denote the horizontal and vertical angular spans of each tile.
Considering heterogenous channel conditions of different users and limited transmission resource, we pre-encode each tile into $L$ representations corresponding to $L$ quality levels using HEVC or H.264, as in \textit{Dynamic Adaptive Streaming over HTTP (DASH)}, and will optimize the quality levels of the tiles that will be transmitted. Let $\mathcal{L}\triangleq\{1,...,L\}$ denote the set of quality levels. For all $l\in\mathcal{L}$, the $l$-th representation of each tile corresponds to the $l$-th lowest quality. For ease of exposition, assume that tiles with the same quality level have the same encoding rate. Let $D_{l}$ denote the encoding rate of the $l$-th representation. We have $D_{1}<D_{2}<...<D_{L}$.

To avoid view switch delay, for each user, the set of tiles that cover its current FoV and the FoVs that may be watched shortly will be delivered.
Let $\Phi_{k}$ denote the set of indices of tiles transmitted to user $k$, and let $\Phi\triangleq\bigcup_{k\in\mathcal{K}}\Phi_{k}$ denote the set of tiles transmitted to users.
In order to make use of multicasting opportunities and avoid redundant transmissions, we divide $\Phi$ into $I$ disjoint non-empty subsets $\mathcal{S}_{i}, i\in\mathcal{I}\triangleq\{1,...,I\}$, where for all $i,j\in \mathcal{I},i\neq j$, $\mathcal{S}_{i}$ and $\mathcal{S}_{j}$ are for different groups of users~\cite{8428401}. Let $\mathcal{K}_{i}$ and $K_{i}=|\mathcal{K}_{i}|$ denote the set and the number of users in the $i$-th group, for all $i\in\mathcal{I}$. When $K_{i}=1$, the server unicasts the tiles in $\mathcal{S}_{i}$ to the single user in $\mathcal{K}_{i}$; when $K_{i}>1$, the server multicasts the tiles in $\mathcal{S}_{i}$ to the users in $\mathcal{K}_{i}$. Without loss of generality, we refer to this transmission as multicast, although both unicast and multicast may happen. Later, we shall see that the tile sets $\mathcal{S}_{i}, i\in\mathcal{I}$ instead of tiles serve as optimization units, enabling complexity reduction compared with~\cite{Ahmadi:2017:AMS:3126686.3126743}, without loss of optimality.

\begin{Exam}
As illustrated in Fig. \ref{system}, consider transmission of a 360 VR video from a single-antenna server to 2 single-antenna users. $K=2$, $M=4$, $N=8$, $\Phi_{1}$$=$$\{(1,3),$$(2,3),$$(1,4),$$(2,4),$$(1,5),$$(2,5)\}$, $\Phi_{2}$$=\{(2,4),$$(3,4),$$(2,5),$$(3,5),$$(2,6),$$(3,6)\}$ and $\Phi=\Phi_{1}\bigcup\Phi_{2}=\{(1,3),$$(2,3),$$(1,4),$$(2,4),$$(1,5),$$(2,5),$$(2,6),$$(3,4),$$(3,5),$$(3,6)\}$. Divide $\Phi$ into $3$ disjoint non-empty subsets $\mathcal{S}_{1}=\{(1,3),(2,3),(1,4),(1,5)\}$, $\mathcal{S}_{2}=\{(2,4),(2,5)\}$ and $\mathcal{S}_{3}=\{(3,4),(3,5),(2,6),(3,6)\}$. $\mathcal{S}_{1}$ is unicasted to $\mathcal{K}_{1}=\{1\}$, $\mathcal{S}_{3}$ is unicasted to $\mathcal{K}_{3}=\{2\}$, and $\mathcal{S}_{2}$ is multicasted to $\mathcal{K}_{2}=\{1,2\}$.
\end{Exam}

We consider time division multiple access (TDMA). Each TDMA frame has a duration of $T$ (in seconds). Consider one frame. The time allocated to transmit the tiles in $\mathcal{S}_{i}$ is denoted by $t_{i}$. Thus, we have the following transmission time allocation constraints:
\begin{align}
t_{i}\geq0,\quad i\in\mathcal{I}, \label{t1}\\
\sum_{i\in\mathcal{I}}t_{i}\leq T.\label{t2}
\end{align}
We consider a narrow band system with bandwidth $B$ (in Hz), and assume block fading, i.e., each channel state (over bandwidth $B$) does not change within one frame. Let $h_{k}$ denote the power of the channel between user $k$ and the server. Let $p_{i}$ denote the transmission power of the symbols for the tiles in $S_i$. We impose the following transmission power allocation constraints:
\begin{align}
p_{i}\geq0,\quad i\in\mathcal{I},\label{p1}\\
\sum_{i\in\mathcal{I}}t_{i}p_{i}\leq Q,\label{p2}
\end{align}
where $Q$ represents the transmission energy limit for one frame.
For all $k\in\mathcal{K}_{i},i\in\mathcal{I}$, the maximum achievable transmission rate (in bit/s) for the tiles in $\mathcal{S}_{i}$ to user $k$ is given by
$B\log_{2}\left(1+\frac{p_{i}h_{k}}{n_{0}}\right)$,
where $n_{0}$ is the power of the complex additive white Gaussian channel noise at each receiver.

Let $x_{m,n}$ denote the selected quality level for tile $(m,n)\in\Phi$, where
\begin{align}
x_{m,n}\in\mathcal{L},\;(m,n)\in\Phi.\label{xl}
\end{align}
Note that $\sum_{(m,n)\in \mathcal{S}_{i}}x_{m,n}$ represents the sum of the selected quality levels of the tiles in $\mathcal{S}_{i}$. Let $\gamma\triangleq\max_{l\in\mathcal{L}}\frac{D_{l}}{l}$ denote the maximum ratio of encoding rate and quality level for all quality levels. Let $h_{i,\min}\triangleq\min_{k\in\mathcal{K}_{i}}h_{k}$ denote the minimum channel power for all users in $\mathcal{K}_{i}$.
To guarantee that all users in $\mathcal{K}_{i}$ can successfully receive the tiles in $\mathcal{S}_{i}$ of the selected quality levels, we require:\footnote{Note that the conservative constraints in \eqref{xtp} are for analytical tractability, and will not lead to much performance degradation, as shown in Sec.~\ref{section_4}.}
\begin{align}
&\gamma T\sum_{(m,n)\in \mathcal{S}_{i}}x_{m,n}\leq t_{i}B\log_{2}\left(1+\frac{p_{i}h_{i,\min}}{n_{0}}\right),\quad i\in\mathcal{I}. \label{xtp}
\end{align}
In addition, $\sum_{(m,n)\in \Phi_{k}}x_{m,n}$ stands for the sum of the selected quality levels of the tiles in $\Phi_{k}$ and can be treated as a measure of the QoE for user $k$. A larger value of $\sum_{(m,n)\in \Phi_{k}}x_{m,n}$ indicates higher QoE for user $k$. Therefore, the total utility of all $K$ users is given by:
\begin{align}
U(\mathbf{x})=\sum_{k\in\mathcal{K}} \sum_{(m,n)\in \Phi_{k}}x_{m,n}.\label{U}
\end{align}

To smooth border effects of mixed-quality tiles, we require that the quality difference between any two adjacent tiles is bounded by a parameter $\Delta\in\mathcal{L}\bigcup\{0\}$. In addition, considering that the first column of tiles are connected to the last column of tiles in a 360 VR video, we have the following smoothness constraints:
\begin{align}
&|x_{m,n}-x_{m,(n+1)\bmod N}|\leq\Delta, \nonumber\\
&\hspace{25mm} (m,n)\in\Phi, \left(m,(n+1)\bmod N\right)\in\Phi,\label{QoE1}\\
&|x_{m,n}-x_{m+1,n}|\leq\Delta, \quad (m,n)\in\Phi, \left((m+1),n\right)\in\Phi.\label{QoE2}
\end{align}
Note that quality smoothness guarantee is not considered in~\cite{Ahmadi:2017:AMS:3126686.3126743}.


\section{Problem Formulation and Suboptimal Solutions}\label{section_3}

In this paper, we would like to optimize the quality selection $\mathbf{x}\triangleq(x_{m,n})_{(m,n)\in\Phi}$, transmission time allocation $\mathbf{t}\triangleq(t_i)_{i\in\mathcal{I}}$ and transmission power allocation $\mathbf{p}\triangleq(p_i)_{i\in\mathcal{I}}$, to maximize the total utility $U(\mathbf x)$ in \eqref{U} subject to the quality selection constraints in \eqref{xl}, \eqref{xtp}, transmission time allocation constraints in \eqref{t1}, \eqref{t2}, transmission power allocation constraints in \eqref{p1}, \eqref{p2}, and quality smoothness constraints in \eqref{QoE1}, \eqref{QoE2}.

%

\begin{problem}[Total Utility Maxmization]\label{P1}
\begin{align}
U^*\triangleq\max_{\mathbf{x},\mathbf{t},\mathbf{p}} \ &\sum_{k\in\mathcal{K}} \sum_{(m,n)\in \Phi_{k}}x_{m,n}\nonumber\\
\text{s.t.} \qquad
&\eqref{t1},\eqref{t2},\eqref{p1},\eqref{p2},\eqref{xl},\eqref{xtp},\eqref{QoE1},\eqref{QoE2}.\nonumber
\end{align}

\end{problem}

Due to the discrete constraints in $\eqref{xl}$, Problem~\ref{P1} is a mixed discrete-continuous optimization problem, which is NP-hard in general. In the following, we propose two low-complexity algorithms to obtain two suboptimal solutions of Problem 1.

\subsection{Suboptimal Solution based on Continuous Relaxation}

In this part, we obtain a suboptimal solution of Problem~\ref{P1} by continuous relaxation. Specifically, by relaxing the discrete constraints in \eqref{xl} to
\begin{align}
x_{m,n}\in[1,L],\quad (m,n)\in\Phi,\label{xcontinue}
\end{align}
we can obtain the continuous relaxation of Problem~\ref{P1}. As the constraint functions in $\eqref{xtp}$ are non-convex, the continuous relaxation of Problem~\ref{P1} is non-convex. In general, we can only obtain a stationary point of a non-convex problem.
By introducing auxiliary variables $e_i=t_i p_i$, $i\in\mathcal{I}$, we can equivalently transform the constraints in \eqref{p1}, \eqref{p2} and \eqref{xtp} to
\begin{align}
&e_{i}\geq0,\quad i\in\mathcal{I},\label{e1}\\
&\sum_{i\in\mathcal{I}}e_{i}\leq Q,\label{e2}\\
&\gamma T\sum_{(m,n)\in \mathcal{S}_{i}}x_{m,n}\leq t_{i}B\log_{2}\left(1+\frac{e_{i}h_{i,min}}{t_{i}n_{0}}\right),\quad i\in\mathcal{I}.\label{xte}
\end{align}
Denote $\mathbf{e}\triangleq(e_i)_{i\in\mathcal{I}}$. Then, we can obtain an equivalent convex formulation of the continuous relaxation of Problem~\ref{P1} as follows.

\begin{problem}[Convex Formulation of Relaxed Problem~\ref{P1}]\label{P2}
\begin{align}
\bar{U}^*\triangleq\max_{\mathbf{x},\mathbf{t},\mathbf{e}} \ &\sum_{k\in\mathcal{K}} \sum_{(m,n)\in \Phi_{k}}x_{m,n}\nonumber\\
\text{s.t.} \qquad
&\eqref{t1},\eqref{t2},\eqref{QoE1},\eqref{QoE2},\eqref{xcontinue},\eqref{e1},\eqref{e2},\eqref{xte}.\nonumber
\end{align}
Let $(\mathbf{x}^*,\mathbf{t}^*,\mathbf{e}^*)$ denote an optimal solution of Problem~\ref{P2}.
\end{problem}

Note that $\bar{U}^*\geq U^*$. An optimal solution of Problem~\ref{P2} can be obtained efficiently using standard convex optimization techniques. But it is usually not in the feasible set of Problem~\ref{P1}, as we have relaxed the discrete constraints in \eqref{xl} of Problem~\ref{P1}. Round down $\mathbf{x}^*$ to ${\lfloor{\mathbf{x}^*}\rfloor}\triangleq({\lfloor{x^*_{m,n}}\rfloor})_{(m,n)\in\Phi}$, where ${\lfloor{x}\rfloor}$ denotes the greatest integer less than or equal to $\mathbf{x}$, and construct $\mathbf{p}^*\triangleq(p_i^*)_{i\in\mathcal{I}}$, where ${p}^*_i\triangleq\frac{e^*_i}{t^*_i}$ for all ${i\in\mathcal{I}}$. We have the following result.

\begin{lemma}[Suboptimal Solution of Problem~\ref{P1}]
$({\lfloor{\mathbf{x}^*}\rfloor},\mathbf{t}^*, \mathbf{p}^*)$ satisfies all the constraints in Problem~\ref{P1}, and $U^*-U({\lfloor{\mathbf{x}^*}\rfloor}) \leq\sum_{k\in\mathcal{K}} \sum_{(m,n)\in \Phi_{k}}({x}^*_{m,n}-{\lfloor{{x}^*_{m,n}}\rfloor})$.
\end{lemma}

Therefore, $({\lfloor{\mathbf{x}^*}\rfloor},\mathbf{t}^*, \mathbf{p}^*)$ can be treated as a suboptimal solution of Problem~\ref{P1}, and an upper bound on its performance gap, i.e., $\sum_{k\in\mathcal{K}} \sum_{(m,n)\in \Phi_{k}}({x}^*_{m,n}-{\lfloor{{x}^*_{m,n}}\rfloor})$, can be easily evaluated. The details of the above procedures are summarized in Algorithm~\ref{alg1}.

\begin{algorithm}
    \caption{Suboptimal Solution of Problem~\ref{P1} based on Continuous Relaxation}
\begin{footnotesize}
	\textbf{Output} $({\lfloor{\mathbf{x}^*}\rfloor},\mathbf{t}^*, \mathbf{p}^*)$.
        \begin{algorithmic}[1]\label{alg1}
           \STATE Compute $(\mathbf{x}^*,\mathbf{t}^*,\mathbf{e}^*)$ of Problem~\ref{P2} using standard convex optimization techniques.\\
           \STATE Set ${\lfloor{\mathbf{x}^*}\rfloor}=({\lfloor{x^*_{m,n}}\rfloor})_{(m,n)\in\Phi}$, and $\mathbf{p}^*=(\frac{e^*_i}{t^*_i})_{i\in\mathcal{I}}$.

    \end{algorithmic}
    \end{footnotesize}
\end{algorithm}

\subsection{Suboptimal Solution based on DC programming}

In this part, we obtain a suboptimal solution by DC programming. First, introduce variables $\mathbf{y}\triangleq (y_{m,n,l})_{(m,n)\in\Phi,l\in\mathcal{L}}$, where

\begin{align}
&y_{m,n,l}\in{\{0,1\}}, \quad (m,n)\in\Phi, l\in\mathcal{L}. \label{y0}
\end{align}
Then, treating $\sum_{l\in\mathcal{L}}y_{m,n,l}$ as $x_{m,n}$ for all $(m,n)\in\Phi$,
the constraints in \eqref{xl} can be equivalently expressed in terms of $\mathbf{y}$ as follows:
\begin{align}
&\sum_{l\in\mathcal{L}}y_{m,n,l}\geq1, \quad (m,n)\in\Phi.\label{ysum}
\end{align}
The discrete constraints in \eqref{y0} can be rewritten as the following continuous constraints:
\begin{align}
&y_{m,n,l}\in[0,1], \quad (m,n)\in\Phi,l\in\mathcal{L},\label{y1}\\
&\sum_{(m,n)\in\Phi}\sum_{l\in\mathcal{L}}y_{m,n,l}(1-y_{m,n,l})\leq 0.\label{y2}
\end{align}

As in Problem~\ref{P2}, we optimize $\mathbf{e}$ instead of $\mathbf{p}$. Then, Problem~\ref{P1} can be equivalently transformed to the following problem.

\begin{problem}[DC Problem of Problem~\ref{P1}]\label{P3}
\begin{align}
U^*=&\max_{\mathbf{y},\mathbf{t},\mathbf{e}} \ \sum_{k\in\mathcal{K}} \sum_{(m,n)\in \Phi_{k}}\sum_{l\in\mathcal{L}}y_{m,n,l}\nonumber\\
&\text{s.t.} \quad \eqref{t1},\eqref{t2},\eqref{QoE1},\eqref{QoE2},\eqref{e1},\eqref{e2},\eqref{xte},\eqref{y1},\eqref{y2},\nonumber
\end{align}
where $x_{m,n}$ in \eqref{QoE1}, \eqref{QoE2}, \eqref{xte} is given by $x_{m,n}=\sum_{l\in\mathcal{L}}y_{m,n,l}$, for all $(m,n)\in\Phi$.
\end{problem}

Note that the objective function of Problem~\ref{P3} is convex, the constraint functions in \eqref{y2} are concave, and the other constraint functions are convex. Thus, Problem~\ref{P3} is a difference of convex (DC) problem \cite{Lipp2016}. In the following, we adopt the DC method in \cite{LeThi2012} to obtain a stationary point of Problem~\ref{P3}. First, we approximate Problem~\ref{P3} by disregarding the constraint in \eqref{y2} and adding to the objective function a penalty for violating the constraint in \eqref{y2}.

\begin{problem}[Penalized Problem of Problem~\ref{P3}]\label{P4}
\begin{align}
&\max_{\mathbf{y},\mathbf{t},\mathbf{e}} \ \sum_{k\in\mathcal{K}} \sum_{(m,n)\in \Phi_{k}}\sum_{l\in\mathcal{L}}y_{m,n,l}-\rho P(\mathbf{y})\nonumber\\
&\text{s.t.} \quad \eqref{t1},\eqref{t2},\eqref{QoE1},\eqref{QoE2},\eqref{e1},\eqref{e2},\eqref{xte},\eqref{y1},\nonumber
\end{align}
where the penalty parameter $\rho>0$ and the penalty function $P(\mathbf{y})$ is given by
\begin{align}
P(\mathbf{y})=\sum_{(m,n)\in\Phi}\sum_{l\in\mathcal{L}}y_{m,n,l}(1-y_{m,n,l}).
\end{align}
\end{problem}

There exists $\rho_0>0$ such that for all $\rho>\rho_0$, Problem ~\ref{P4} is equivalent to Problem~\ref{P3} \cite{LeThi2012}. Now, we solve Problem~\ref{P4} instead of Problem~\ref{P3} by using a DC algorithm.
Specifically, at the $j$-th iteration, we update the solution $(\mathbf{y}^{(j)},\mathbf{t}^{(j)},\mathbf{e}^{(j)})$ by solving the following approximate problem.

\begin{problem}\label{P5}\textit{(Convex Approximation of Problem~\ref{P4} at $j$-th}
\textit{Iteration):}
\begin{align}
(\mathbf{y}^{(j)},\mathbf{t}^{(j)},&\mathbf{e}^{(j)})\triangleq\nonumber\\
&\mathop{\arg\max}_{\mathbf{y},\mathbf{t},\mathbf{e}} \ \sum_{k\in\mathcal{K}} \sum_{(m,n)\in \Phi_{k}}\sum_{l\in\mathcal{L}}y_{m,n,l}-\rho \hat{P}(\mathbf{y};\mathbf{y}^{(j-1)})\nonumber\\
&\text{s.t.} \quad \eqref{t1},\eqref{t2},\eqref{QoE1},\eqref{QoE2},\eqref{e1},\eqref{e2},\eqref{xte},\eqref{y1},\nonumber
\end{align}
where
\begin{align}
&\hat{P}(\mathbf{y};\mathbf{y}^{(j-1)})\triangleq P(\mathbf{y}^{(j-1)})+\nabla P(\mathbf{y}^{(j-1)})^T (\mathbf{y}-\mathbf{y}^{(j-1)})\nonumber\\
&=\sum_{(m,n)\in\Phi}\sum_{l\in\mathcal{L}} \left(\big(1-2y_{m,n,l}^{(j-1)}\big)y_{m,n,l}+\big(y_{m,n,l}^{(j-1)}\big)^2\right).\nonumber
\end{align}
Here, $\mathbf{y}^{(j-1)}\triangleq \left(y_{m,n,l}^{(j-1)}\right)_{(m,n)\in\Phi,l\in\mathcal{L}}$ denotes the solution of Problem~\ref{P5} at the $(j-1)$-th iteration.
\end{problem}

It has been shown that the DC algorithm can obtain a stationary point of Problem~\ref{P4}, denoted by $(\mathbf{y}^{\star},\mathbf{t}^{\star},\mathbf{e}^{\star})$. Due to the equivalence among Problems~\ref{P1}, \ref{P3} and \ref{P4}, $\left(\mathbf{x}^{\star},\mathbf{t}^{\star},\mathbf{p}^{\star}\right)$ can be treated as a suboptimal solution of Problem~\ref{P1}, where $\mathbf{x}^{\star}\triangleq({x}_{m,n}^{\star})_{(m,n)\in\Phi}$ with ${x}_{m,n}^{\star}=\sum_{l\in\mathcal{L}}y_{m,n,l}^{\star}$ for all ${(m,n)\in\Phi}$, and $\mathbf{p}^{\star}\triangleq(p_i)_{i\in\mathcal{I}}$ with ${p}_i^{\star}=\frac{e_i^{\star}}{t_i^{\star}}$ for all ${i\in\mathcal{I}}$. The details are summarized in Algorithm~\ref{alg2}.

\begin{figure*}[!ht]
\normalsize{
\centering
\subfigure[  Total utility vs bandwidth at $Q=0.05$ J and $T=0.05$ s.]{ 
\begin{minipage}{6.25cm}
\centering
\includegraphics[width=6.25cm]{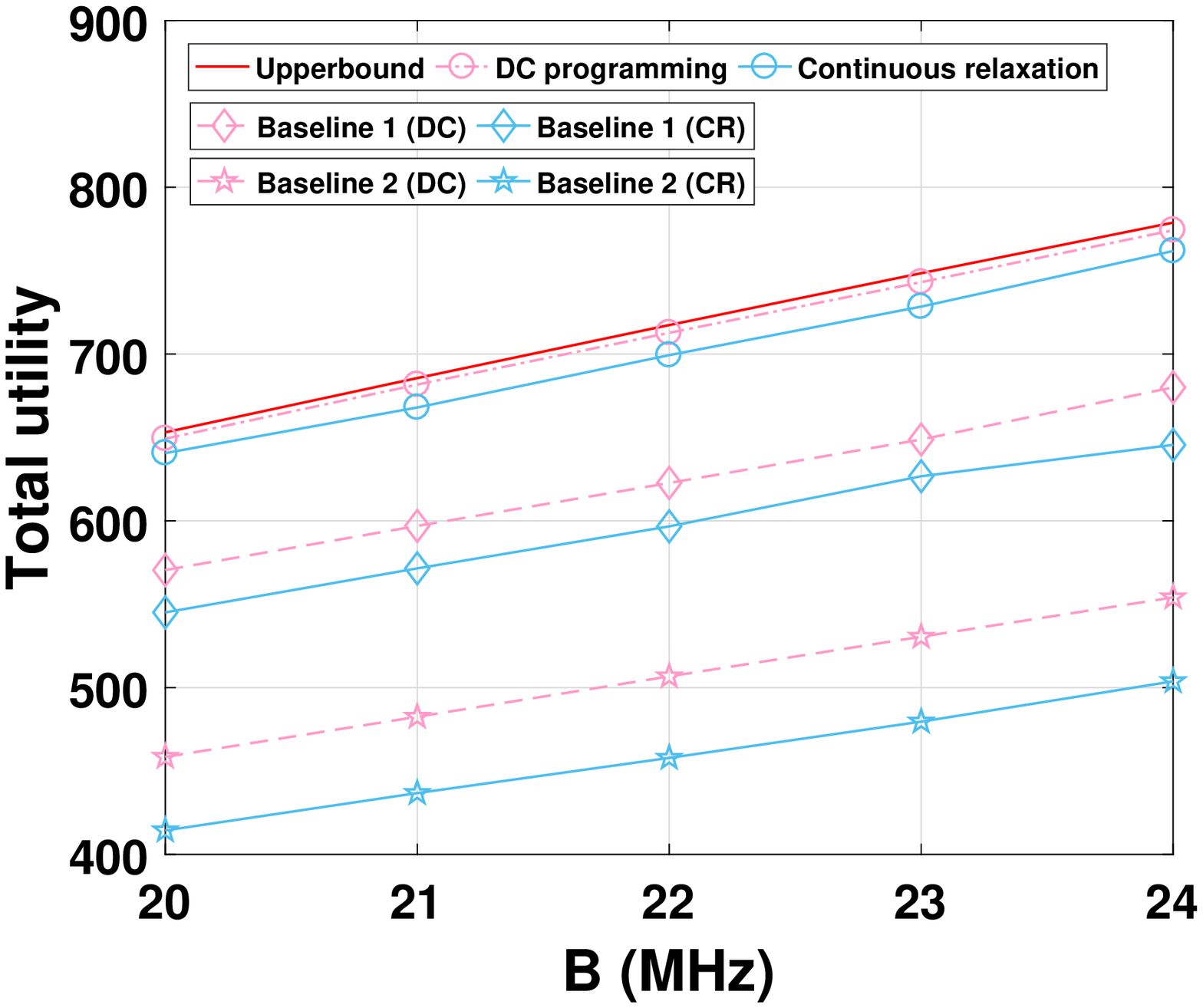}
\end{minipage}
}
\subfigure[ Total utility vs transmission energy limit at $B=20$ MHz and $T=0.05$ s. ]{ 
\begin{minipage}{6.25cm}
\centering
\includegraphics[width=6.25cm]{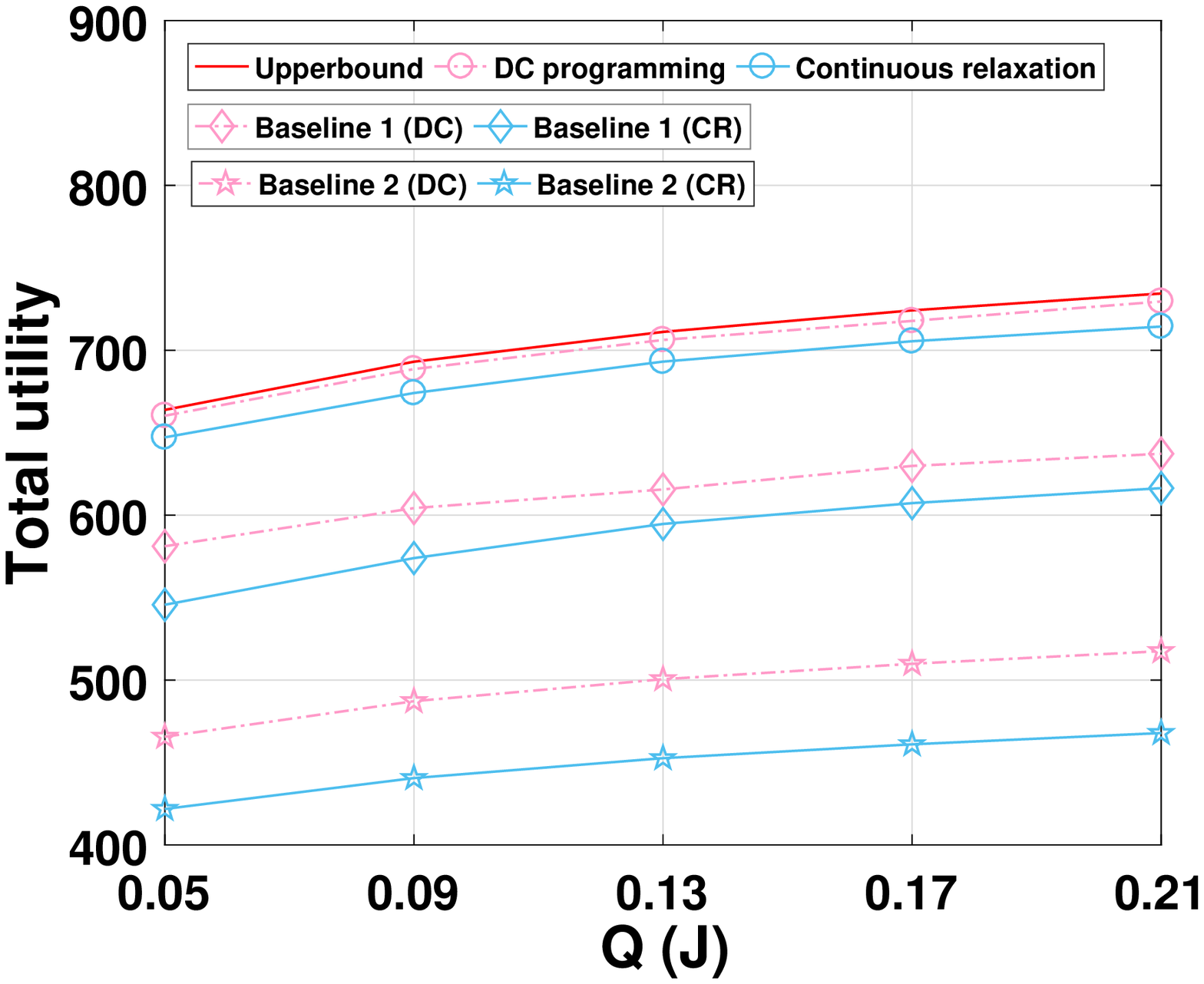}
\end{minipage}
}
\subfigure[ Total utility vs frame duration at $Q=0.05$ J and $B=20$ MHz. ]{ 
\begin{minipage}{6.25cm}
\centering
\includegraphics[width=6.25cm]{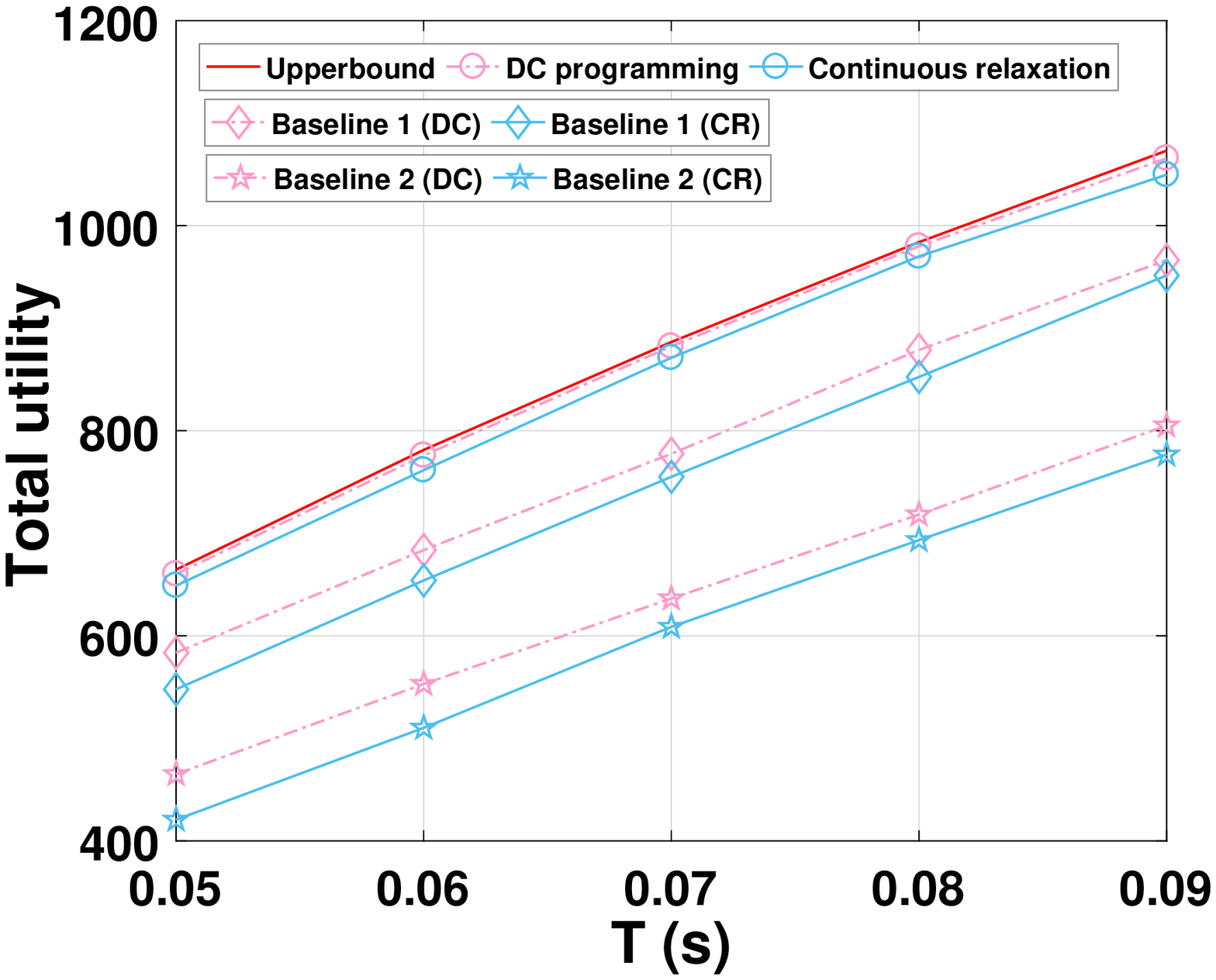}
\end{minipage}
}
\subfigure[ Average PSNR vs bandwidth at $Q=0.05$ J and $T=0.05$ s. ]{ 
\begin{minipage}{6.25cm}
\centering
\includegraphics[width=6.25cm]{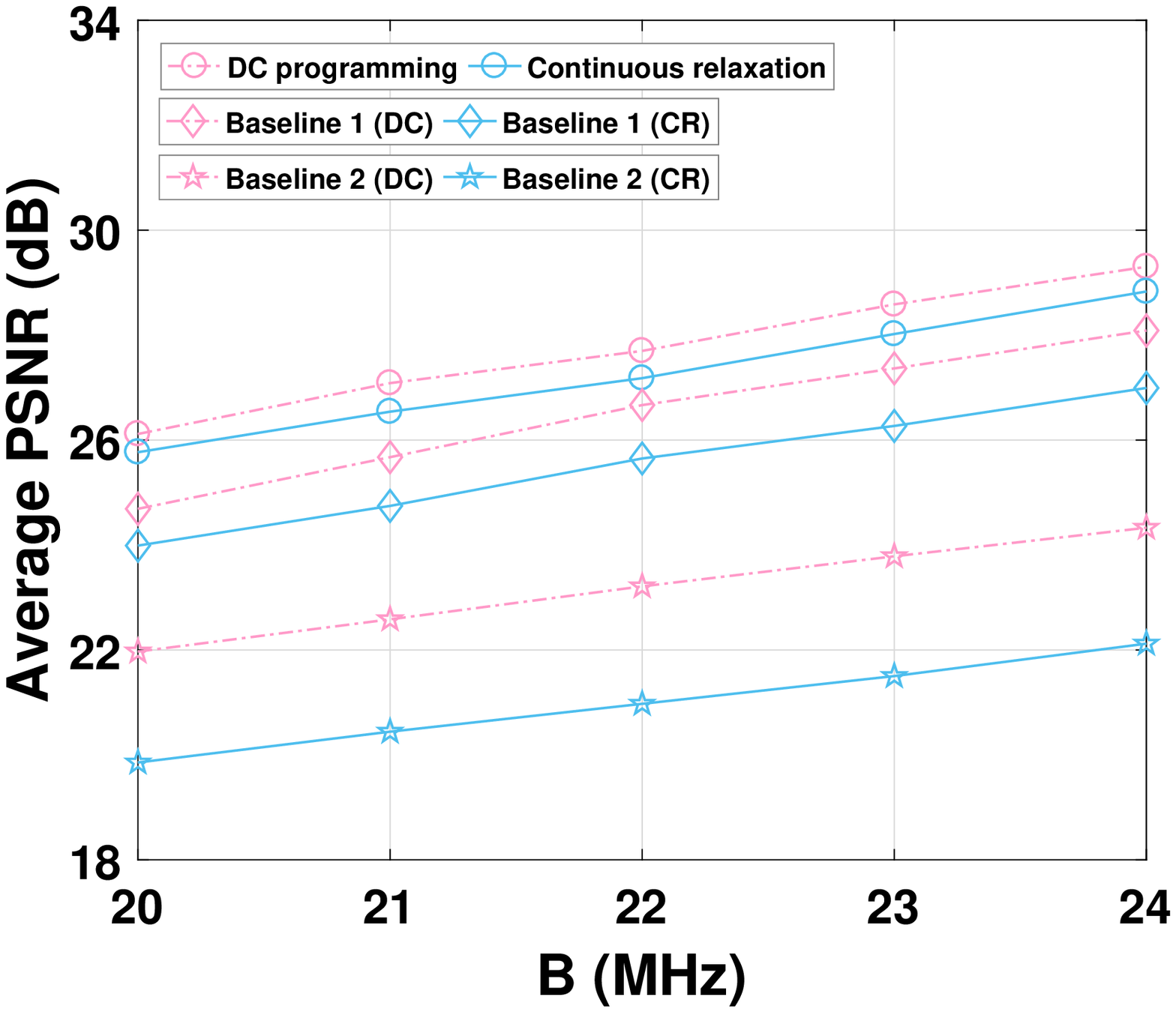}
\end{minipage}
}
\subfigure[ Average PSNR vs transmission energy limit at $B=20$ MHz and $T=0.05$ s. ]{ 
\begin{minipage}{6.25cm}
\centering
\includegraphics[width=6.25cm]{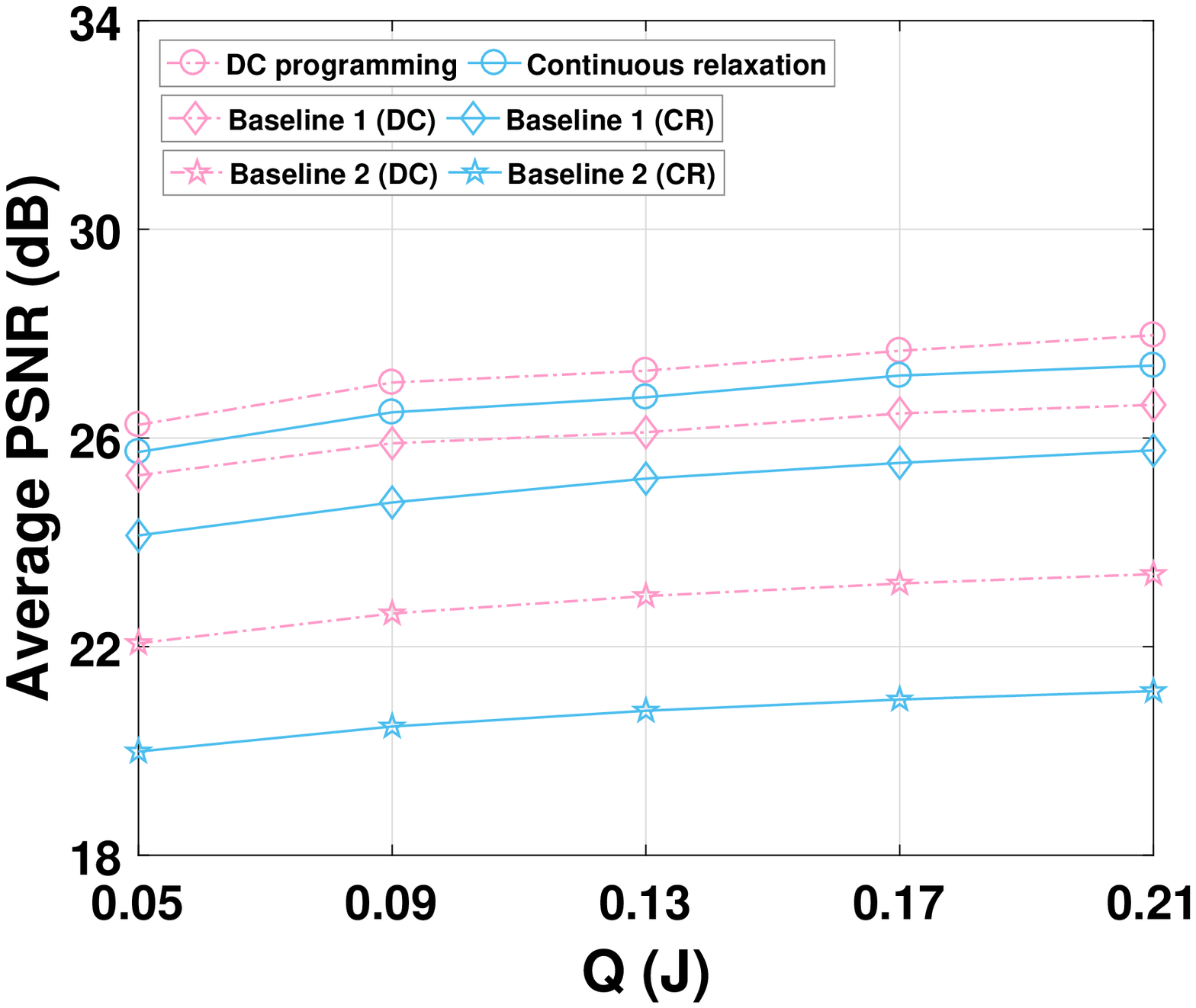}
\end{minipage}
}
\subfigure[ Average PSNR vs frame duration at $Q=0.05$ J and $B=20$ MHz. ]{ 
\begin{minipage}{6.25cm}
\centering
\includegraphics[width=6.25cm]{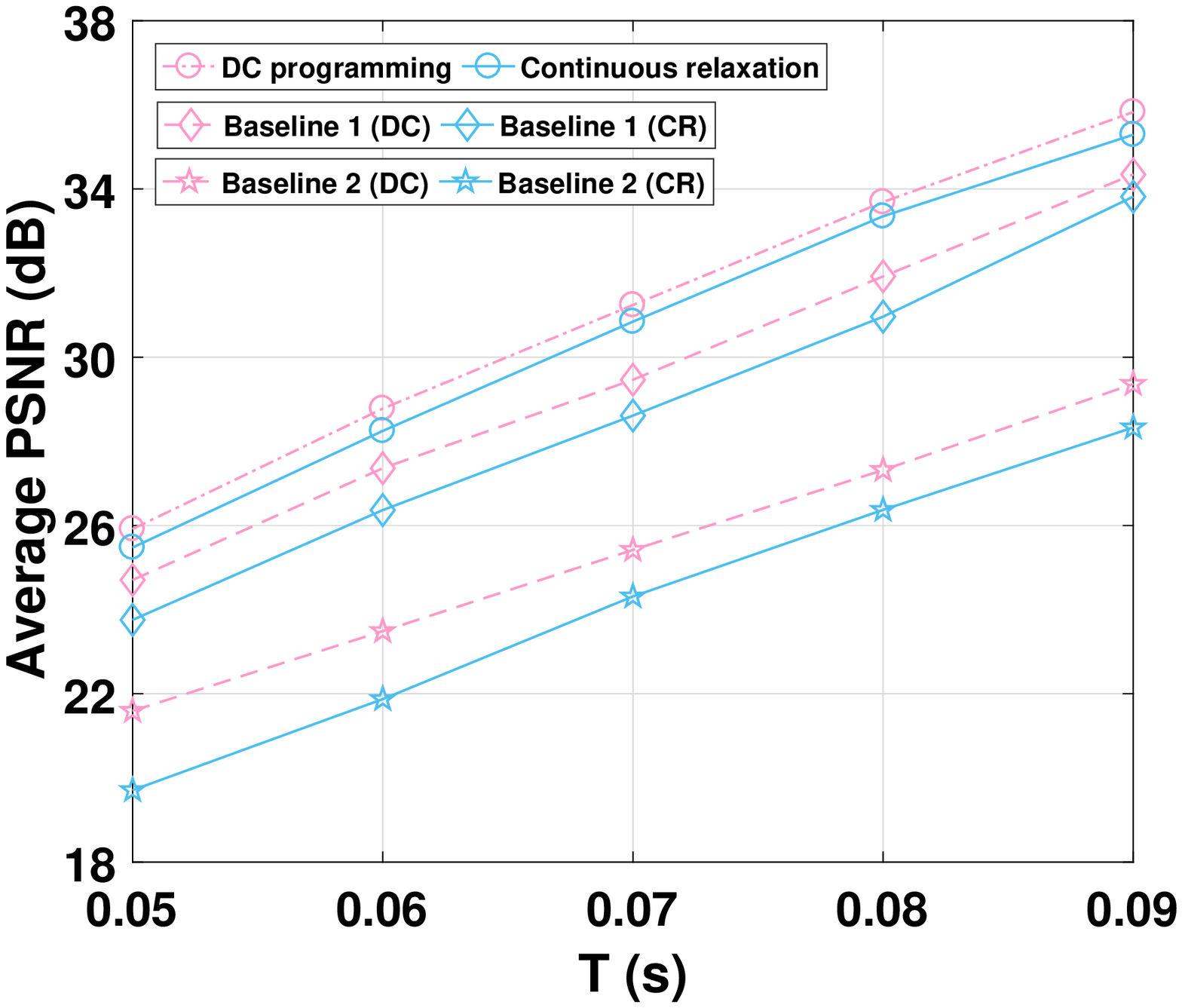}
\end{minipage}
}
\caption{ Comparison between the proposed suboptimal solutions and four baseline schemes. }
\label{example2}
}
\end{figure*}

\begin{algorithm}
\caption{Suboptimal Solution of Problem~\ref{P1} based on DC Programming}
\begin{footnotesize}
\textbf{Output} ($\mathbf{x}^\star$,$\mathbf{t}^\star$,$\mathbf{p}^\star$).
\begin{algorithmic}[1]\label{alg2}
\STATE \textbf{Initial Step}: Find an initial feasible point $(\mathbf{y}^{(0)},\mathbf{t}^{(0)},\mathbf{e}^{(0)})$ of Problem~\ref{P4}, choose a sufficiently large $\rho$, and set $j=0$.\\
\STATE \textbf{repeat}
\STATE Set $j=j+1$.
\STATE Obtain $(\mathbf{y}^{(j)},\mathbf{t}^{(j)},\mathbf{e}^{(j)})$ of Problem~\ref{P5} using standard convex optimization techniques.
\STATE \textbf{until} convergence criteria is met.
\STATE Set $(\mathbf{y}^{\star},\mathbf{t}^{\star},\mathbf{e}^{\star})=(\mathbf{y}^{(j)},\mathbf{t}^{(j)},\mathbf{e}^{(j)})$.
\STATE Set $\mathbf{x}^{\star}=(\sum_{l\in\mathcal{L}}y_{m,n,l}^{\star})_{(m,n)\in\Phi}$, and $\mathbf{p}^{\star}=(\frac{e_i^{\star}}{t_i^{\star}})_{i\in\mathcal{I}}$.

\end{algorithmic}
\end{footnotesize}
\end{algorithm}

\section{Simulation}\label{section_4}
In the simulation, we consider the following setting. Consider $K=2$ and both path loss and small-scale Rayleigh fading. For all $k\in \mathcal{K}$, assume channel power $h_{k}$ follows the exponential distribution with mean $10^{-3}$ (which is to reflect the path loss). The complex additive white Gaussian channel noise is $n_{0}=Bk_{B}T_{0}$, where $k_{B}=1.38\times10^{-23}$ Joule/Kelvin is the Boltzmann constant and $T_{0}=300$ Kelvin is the temperature. Consider $\Delta=1$. We set the horizontal and vertical angular spans of each FoV as $F_{h}=F_{v}=100^{\circ}$\cite{Ju:2017:UWV:3097895.3097899}. Considering possible view changes in a short period, besides each FoV we transmit an extra $10^{\circ}$ in every direction. Consider $M\times N=18\times36$, $\Phi_{1}=\{(m,n)\mid 2\leq m\leq13,m\in\mathcal{M},10\leq n\leq21,n\in\mathcal{N}\}$ and $\Phi_{2}=\{(m,n)\mid 7\leq m\leq18,m\in\mathcal{M},15\leq n\leq26,n\in\mathcal{N}\}$. We use 360 video sequence \emph{Reframe Iran}  from YouTube as the video source and divide it into 
tiles with $L=6$ quality levels. The parameters are shown in TABLE \ref{2}. The encoding is done with the HEVC codec Kvazaar.  From TABLE \ref{2}, we know $\gamma\triangleq\max_{l\in\mathcal{L}}\frac{D_{l}}{l}=8.408\times10^5$. Besides the total utility defined in the paper, we also consider the average PSNR for each FoV as the performance measure. We generate $100$ random independent channel realizations, and evaluate the average performance over these realizations. We use Matlab software and cvx tool box to implement Algorithm~\ref{alg1} and Algorithm~\ref{alg2}.

\begin{table}[h]
\centering
\caption{\label{2}Encoding rate and PSNR of a tile at different quality levels. }
\setlength{\tabcolsep}{1.6mm}{
\begin{tabular}{|c|c|c|c|c|c|c|}
\hline
Quality level &1&2&3&4&5&6\\
\hline
Quantization parameter &42&35&28&21&14&7\\
\hline
Encoding rate ($\times10^{5}$) &6.66&16.18&24.29&32.01&40.23&50.45\\
\hline
PSNR &15.82&25.24&32.86&39.96&46.11&50.96\\
\hline
\end{tabular}}
\end{table}


In the simulation, we compare two proposed suboptimal solutions with four baseline schemes. Baseline 1 (CR) and Baseline 1 (DC) both consider multicast opportunities and quality level optimization but with equal transmission time and power allocation, i.e., $t_{i}=\frac{T}{I}$ and $p_{i}=\frac{|\mathcal{S}_{i}|}{\sum_{i}|\mathcal{S}_{i}|}Q\; $for all $i\in\mathcal{I}$, where $|\mathcal{S}_i|$ denotes the number of tiles in $\mathcal{S}_i$.
Given the equal transmission time and power allocation, Baseline 1 (CR) corresponds to a suboptimal quality level selection based on continuous relaxation  (similar to Algorithm~\ref{alg1}) and Baseline 2 (DC) corresponds to a suboptimal quality level selection based on DC programming (similar to Algorithm~\ref{alg2}).
Baseline 2 (CR) and Baseline 2 (DC) both consider optimal quality level selection, as well as transmission time and power allocation but without exploiting multicast opportunities, i.e., different users are served separately no matter whether $\Phi_k,\; k \in\mathcal{K}$ are disjoint or not. The optimization problem for unicast can be formulated in a similar way to Problem 1. Baseline 2 (CR) corresponds to a suboptimal solution of this new problem based on continuous relaxation (similar to Algorithm~\ref{alg1}), and Baseline 2 (DC) corresponds to a suboptimal solution based on DC programming (similar to Algorithm~\ref{alg2}).
The benchmark \textit{Upperbound} corresponds to the optimal value of Problem ~\ref{P2}, i.e., $\bar{U}^*$.


Fig. \ref{example2} illustrates the total utility and the average PSNR (for each FoV) versus the bandwidth $B$, the transmission energy limit $Q$ and the frame duration $T$, respectively. From Fig. \ref{example2}, we can see that the total utility and the average PSNR of each scheme both increase with $B$, $Q$ and $T$. In addition, we can see that each scheme based on DC programming outperforms the corresponding scheme based on continuous relaxation, due to the performance loss caused by rounding in the continuous relaxation approach. As illustrated in TABLE \ref{1}, the continuous relaxation approach has a lower complexity than the DC programming approach. Finally, we can see that the two proposed suboptimal solutions outperform all the baselines and achieve near optimal performance (as their performances are close to the upperbound of the optimal performance), demonstrating the effectiveness of the suboptimal solutions.

\begin{table}[h]
\centering
\caption{\label{1}Computation times and total utilities of the two suboptimal solutions at $Q=0.05$ J, $B=20$ MHz and $T=0.05$ s.}
\begin{tabular}{|c|c|c|}
\hline
          & Computation time (s) & total utility \\
\hline
Continuous relaxation &  9.84 & 527.76\\
\hline
DC programming & 282.98 & 534.61 \\
\hline
\end{tabular}
\end{table}

\section{Conclusions}\label{section_5}

In this paper, we consider multi-quality multicast of a 360 VR video from a single server to multiple users using TDMA. To improve transmission efficiency, tiling is adopted, and each tile is pre-encoded into multiple representations with different qualities. Sets of tiles that are transmitted to different groups of users are considered to enable complexity reduction without out loss of optimality. We optimize the quality level selection, transmission time allocation and transmission power allocation for the tile sets to maximize the total utility of all users under the transmission time and power allocation constraints as well as the quality smoothness constraints for mixed-quality tiles. The problem is a challenging mixed discrete-continuous optimization problem. We propose two low-complexity algorithms to obtain two suboptimal solutions, using continuous relaxation and DC programming, respectively. Finally, numerical results demonstrate the advantage of the proposed suboptimal solutions. To the best of our knowledge, this is the first work providing low-complexity optimization-based multi-quality multicast for 360 VR video with quality smoothness guarantee.

\end{document}